\documentclass[apjl]{emulateapj}
\usepackage{apjfonts}
\usepackage{graphicx}

\newcommand{\lya}{\mbox{${\rm Ly}\alpha$}}

\newcommand{\apll}{\lesssim}
\newcommand{\etal}{\ensuremath{\mbox{et~al.}}}

\shorttitle{Dusty Galaxy Hosting a Dark GRB at $z=3$}

\shortauthors{Chen \etal}

\begin{document}

\slugcomment{Accepted for Publication in the Astrophysical Journal Letters}

\title{A Mature Dusty Star-forming Galaxy Hosting GRB\,080607 at $z=3.036^1$}

\author{HSIAO-WEN CHEN\altaffilmark{2}, 
DANIEL A.\ PERLEY\altaffilmark{3},
CHRISTINE D.\ WILSON\altaffilmark{4},
S.\ BRADLEY CENKO\altaffilmark{3},
ANDREW J.\ LEVAN\altaffilmark{5},
JOSHUA S.\ BLOOM\altaffilmark{3},
JASON X.\ PROCHASKA\altaffilmark{6},
NIAL R.\ TANVIR\altaffilmark{7}
MIROSLAVA DESSAUGES-ZAVADSKY\altaffilmark{8},
MAX PETTINI\altaffilmark{9},
}

\altaffiltext{1}{Based in part on observations made with the NASA/ESA
Hubble Space Telescope, obtained at the Space Telescope Science
Institute, which is operated by the Association of Universities for
Research in Astronomy, Inc., under NASA contract NAS 5-26555. }
\altaffiltext{2}{Dept.\ of Astronomy \& Astrophysics and Kavli
Institute for Cosmological Physics, University of Chicago, Chicago,
IL, 60637, U.S.A.  {\tt hchen@oddjob.uchicago.edu}}
\altaffiltext{3}{Department of Astronomy, 601 Campbell Hall,
University of California, Berkeley, CA 94720 }
\altaffiltext{4}{Department of Physics \& Astronomy, McMaster
University, Hamilton, Ontario, L8S 4M1, Canada}
\altaffiltext{5}{Department of Physics, University of Warwick, Coventry CV4 7AL, UK}
\altaffiltext{6}{UCO/Lick Observatory; University of California, Santa
Cruz, Santa Cruz, CA 95064}
\altaffiltext{7}{Department of Physics and Astronomy, University of Leicester, University Road, Leicester LE1 7RH, UK}
\altaffiltext{8}{Observatoire de Gen\`eve, 51 Ch. des Maillettes,
        1290 Sauverny, Switzerland} 
\altaffiltext{9}{Institute of Astronomy, Madingley Rd., Cambridge, CB3 0HA, UK}

\begin{abstract}

  We report the discovery of the host galaxy of dark burst GRB\,080607
  at $z_{\rm GRB}=3.036$.  GRB\,080607 is a unique case of a highly
  extinguished ($A_V \approx 3$ mag) afterglow that was yet
  sufficiently bright for high-quality absorption-line spectroscopy.
  The host galaxy is clearly resolved in deep HST WF3/IR F160W images
  and well detected in the Spitzer IRAC 3.5$\mu$m and 4.5$\mu$m
  channels, while displaying little/no fluxes in deep optical images
  from Keck and Magellan.  The extremely red optical--infrared colors
  are consistent with the large extinction seen in the afterglow
  light, suggesting that the large amount of dust and gas surface mass
  density seen along the afterglow sightline is not merely local but
  likely reflects the global dust content across the entire host
  galaxy.  Adopting the dust properties and metallicity of the host
  ISM derived from studies of early-time afterglow light and
  absorption-line spectroscopy, we perform a stellar population
  synthesis analysis of the observed spectral energy distribution to
  constrain the intrinsic luminosity and stellar population of this
  dark burst host.  The host galaxy is best described by an
  exponentially declining star formation rate of e-folding time
  $\tau=2$ Gyr and an age of $\sim 2$ Gyr.  We also derive an
  extinction corrected star formation rate of ${\rm SFR}\approx
  125\,h^{-2}\,{\rm M}_\odot\,{\rm yr}^{-1}$ and a total stellar mass
  of $M_*\sim 4\times 10^{11}\ h^{-2}\,{\rm M}_\odot$.  Our study
  provides an example of massive, dusty star-forming galaxies
  contributing to the GRB host galaxy population, supporting the
  notion that long-duration GRBs trace the bulk of cosmic star
  formation.

\end{abstract}


\section{INTRODUCTION}

Early-time spectra of bright $\gamma$-ray burst (GRB) afterglows have
revealed numerous absorption features that allow accurate measurements
of the chemical composition, dust content, and kinematics in the
interstellar medium (ISM) of the host galaxies (e.g.\ Fynbo \etal\
2006; Savaglio 2006; Prochaska \etal\ 2007; 2008).
But as much as $50$\% of long-duration GRBs show a significant
suppression in their optical afterglow light (Jakobsson \etal\ 2004;
Cenko \etal\ 2009).  While some of these "dark" bursts occur during
the reionization epoch at redshifts $z>6$ (e.g.\ Kawai \etal\ 2006;
Greiner \etal\ 2009; Tanvir \etal\ 2009; Salvaterra \etal\ 2009), most
result from large extinction columns in the ISM surrounding massive
star-forming regions at more typical redshifts of $z=1-4$ (e.g.\
Perley et al.\ 2009).

GRB\,080607 at redshift $z_{\rm GRB} = 3.0363$ is a unique case of a
highly extinguished ($A_V \approx 3$ mag) afterglow that was yet
sufficiently bright for high-quality spectroscopy (Prochaska \etal\
2009).  The afterglow spectrum displays positive detections of CO
$A-X$ bandheads (Morton \& Noreau 1994)
that have also
been seen through translucent molecular gas of the Milky Way (e.g.\
Sonnentrucker \etal\ 2007). The presence of Ge\,II 1602 and O\,I 1355
absorption features indicates that the host ISM has been enriched to
roughly solar metallicity.  Identifications of vibrationally excited
H$_2$ indicate the presence of substantial molecular gas at a few
hundred pc from the burst (Sheffer \etal\ 2009).  The large gas
surface mass density ($\approx 400\,{\rm M}_\odot\, {\rm pc}^{-2}$)
and large molecular gas content found in GRB\,080607 are unprecedented
among either damped \lya\ absorbers along random QSO sightlines or GRB
host galaxies (c.f.\ Srianand \etal\ 2008; Noterdaeme \etal\ 2009)
Contrary to the common expectation of GRBs occurring preferentially in
low-mass and low-metallicity environments (e.g.\ Fruchter \etal\
2006), the observed large metal and dust contents, together with the
mass-metallicity relation known for $z=2-3$ galaxies (e.g.\ Mannucci
\etal\ 2009), imply that the host galaxy is massive and intrinsically
luminous.  Searching for the host galaxy of GRB\,080607 therefore
bears significantly on our general understanding of GRB host galaxies
and particularly the dark burst population.

Here we report the discovery of the host galaxy of GRB\,080607 in an
extensive imaging follow-up campaign.  The observed broad-band
spectral energy distribution (SED) from optical to IR wavelengths,
together with known dust properties from studies of early-time
afterglow light (Perley \etal\ 2010) and known ISM metallicity from
absorption-line spectroscopy (Prochaska \etal\ 2009), allow us to
constrain the intrinsic luminosity and stellar population of the host
galaxy of this dark burst.  We adopt a $\Lambda$ cosmology,
$\Omega_{\rm M}=0.3$ and $\Omega_\Lambda = 0.7$, with a dimensionless
Hubble constant $h = H_0/(100 \ {\rm km} \ {\rm s}^{-1}\ {\rm
Mpc}^{-1})$ throughout the paper.

\section{MULTI-WAVELENGTH IMAGING OBSERVATIONS AND DATA ANALYSIS}

\begin{figure*}
\begin{center}
\includegraphics[scale=0.6]{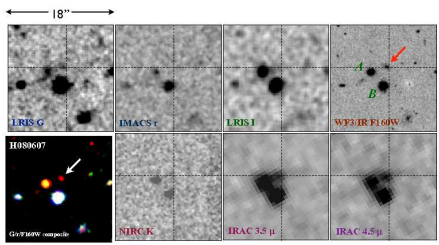}
\caption{Astrometrically aligned images, $18''$ on a side of the field
  surrounding GRB\,080607 at $z_{\rm GRB}=3.036$.  North is up and
  East to the left.  Astrometric solutions were obtained using known
  SDSS objects with a mean r.m.s.\ scatter of $\approx 0.1''$.  The
  position of the early-time afterglow is at the cross with $0.05''$
  error.  The ground-based images have been smoothed over the FWHM of
  the PSF for revealing faint features.  An extended source is clearly
  detected at $0.35''$ away from the afterglow position in the HST
  WF3/IR F160W image (arrow in the top right panel), which is also
  well detected in the Spitzer IRAC 3.5$\mu$m and 4.5$\mu$m images
  (two bottom right panels).  While the souce is not seen in the
  ground-based $r$ and $I$ images, it displays traces of flux in the
  deep $G$- and $K$-band images.  The consistent red color
  (bottom-left panel) and astrometric match with the afterglow lead us
  to conclude that this extended source is the most likely host
  galaxy, H\,080607.  Sources $A$ and $B$ exhibit bluer colors and are
  likely the absorbing galaxies of two strong Mg\,II absorbers
  (Prochaska \etal\ 2009) at $z=1.462$ and $z=1.361$, respectively. }
\end{center}
\end{figure*}

We have carried out an extensive search of the host galaxy of
GRB\,080607 in ground-based and space-based imaging observations.
Deep optical images of the field around GRB\,080607 were obtained
using the Low Resolution Imaging Spectrometer (LRIS; Oke et al.\ 1995)
on the 10~m Keck I telescope on the night of 2009-Feb-19 UT.  Total
integration times of 2490 s and 2220 s were taken through the $G$- and
$I$-band, respectively.  Deep optical $r$-band images were obtained
using the short camera in the IMACS multi-object imaging spectrograph
(Dressler \etal\ 2006) on the Magellan Baade telescope on the night of
2009-May-22 UT.  Three exposures of 600 s each were acquired.  Images were
flat-fielded, registered, and stacked using standard techniques.  The
final stacked images have a mean seeing of $0.8''$, $0.8''$, and
$1.1''$ in $G$, $r$, and $I$, respectively, and are calibrated using
common stars in the SDSS photometric catalog.

Near-IR images of the field were obtained using the Near Infrared
Camera (NIRC) and the $K_s$ filter on Keck I, on the night of
2009-May-31 UT.  A total of 36 exposures of 100 s each 
were acquired.  Images were processed and stacked using standard
techniques via a custom Python pipeline.  The stacked image has a mean
seeing of $0.75''$ and is calibrated using standard stars observed
throughout the night.


Deep near-infrared images of the field were obtained using the IR
channel in Wide Field Camera 3 (WFC3) and the F160W filter on board
the Hubble Space Telescope (PI: Chen).  The observations were carried
out on 2010-Jul-25 UT. 
Three sets of four exposures (900 s each) were obtained.  Individual
exposures were reduced using standard pipeline techniques, corrected
for geometric distortion using drizzle, registered to a common origin,
filtered for deviant pixels,
and combined to form a final stacked image.

Infrared images of the field were also obtained on 2010-Aug-08 UT by
the Spitzer Space Telescope (PI: Perley) in both available
warm-mission IRAC channels (3.6 and 4.5 $\mu$m).  A total of 45
exposures of 100 s each were acquired in each channel using a cycling
dither pattern.  The post-BCD calibrated mosaics were retrieved from
the Spitzer data archive.  Finally, we obtained 850~$\mu$m images of
the GRB field under the SCUBA-2 (Holland \etal\ 2006) Shared Risk
Observations on the James Clerk Maxwell Telescope (M09BI109, PI:
Wilson) on 2010-Mar-14 and 2010-Mar-24.  A total integration time of
two hours was acquired.  The data were reduced and calibrated as
described in Dempsey \etal\ (2010).  The combined image has an r.m.s.\
noise of 2.8 mJy per beam with a beam size of $15''$.

\begin{deluxetable}{lcr}
\tablecaption{\sc{Summary of GRB\,080607 Host Photometry}}
\tablewidth{0pt}
\tablehead{\colhead{Telescope/Instrument} & \colhead{Bandpass} & \colhead{Brightness\tablenotemark{a}}}
\startdata
Keck/LRIS      & $G$       & $AB=28.2\pm 0.4$ \\ 
Magellan/IMACS & $r$       & $AB>27.0$        \\ 
Keck/LRIS      & $I$       & $AB>27.0$        \\ 
HST/WF3/IR     & F160W     & $AB=24.72\pm 0.03$  \\ 
Keck/NIRC      & $Ks$      & $AB=24.8\pm 0.7$    \\ 
Spitzer/IRAC   & 3.5\,$\mu$m & $AB=20.5\pm 0.1$ \\ 
Spitzer/IRAC   & 4.5\,$\mu$m & $AB=19.9\pm 0.1$ \\ 
JCMT/SCUBA-2   & 850\,$\mu$m & $f<7.6$ mJy      \\ 
\enddata
\tablenotetext{a}{When the host is not detected, we measure a
2-$\sigma$ upper limit for the optical and near-IR bandpasses and a
4-$\sigma$ limit for the 850\,$\mu$m flux.  All magnitudes are
corrected for Galactic extinction, $E(B-V)=0.023$.}
\end{deluxetable}


Optical and IR images of the field surrounding GRB\,080607 are
presented in Figure 1.  Astrometric solutions were obtained using
known SDSS objects with a mean r.m.s.\ scatter of $\approx 0.1''$.
The location of the GRB afterglow is marked by the cross, which is
determined based on relative astrometry using seven common stars in
our HST observations and in early-time afterglow images obtained $\sim
2$ hours post burst by UKIRT.  The relative astrometry provides a
precise afterglow position with an error radius of $0.05''$ (c.f.\
Mangano \etal\ 2008).  At $0.35''$ away, an extended source is clearly
detected in the WF3/IR F160W image with a half-light radius of
$r_{1/2}\approx 0.3''$.  We measure $AB({\rm F160W})=24.72\pm 0.03$
over a $0.8''$-radius aperture for this source.

To determine whether this source is the host galaxy of GRB\,080607, we
estimate the probability of finding a random foreground galaxy that
has its optical disk intercepting the afterglow sightline
At close projected distances ($\apll 20\ h^{-1}$ kpc), we expect that
any foreground galaxy would imprint a strong Mg\,II absorption feature
in the afterglow spectrum (e.g.\ Chen \etal\ 2010).  The available
afterglow spectrum of GRB\,080607 allows observations of Mg\,II
absorbers at $z=0.895-2.2$, and indeed two strong Mg\,II absorbers
have been detected at $z=1.341$ and $z=1.462$ (Prochaska \etal\ 2009).
At the same time, galaxies $A$ and $B$ are seen within $3.5''$ radius
of the afterglow sightline with $AB({\rm F160W})=22.4$ and $AB({\rm
F160W})=22.0$, respectively.  Their observed optical and near-IR
colors are bluer than the host candidate (bottom-left panel of Figure
1) and are consistent with galaxies at $z\sim 1.4$.  The projected
distances are within the expected extent of strong Mg\,II absorbers
(Chen \etal\ 2010).  We therefore attribute the two strong Mg\,II
absorbers to galaxies $A$ and $B$.
%
To estimate the probability of a random galaxy at $z<0.895$ occuring
within $2\times r_{1/2}$ of the afterglow position, we calculate the
volume density of galaxies with $L>0.025\,L_*$ (corresponding to
$AB({\rm F160W})=24.7$ at $z=0.895$) using the galaxy luminosity
function of Faber \etal\ (2007).  We find that the probability of
finding a random $z<0.895$ galaxy within this small volume is $<1$\%
which,
together with the presence of $A$ and $B$, lead us to conclude that
the extended source at the afterglow position is the most likely host
of GRB\,080607.

The host is also detected in the $G$-band image with $AB(g)=28.2\pm
0.4$ over a $0.8''$-radius aperture, and in the IRAC 3.5$\mu$m and
4.5$\mu$m channels with $AB({\rm 3.5}\mu)=20.5\pm 0.1$ and $AB({\rm
  4.5}\mu)=19.9\pm 0.1$ over a $1.2''$-radius aperture.  The adopted
aperture sizes roughly matches the size of the apertures adopted for
the optical and near-IR images after accounting for the differences in
the PSFs.  The errors in the IRAC photometry of the host are dominated
by contaminating light from $A$ and $B$, and are estimated based on
the summed flux in a sequence of increasing-diameter apertures.  In
the $K$-band image, we observe at the host position a 1.5-$\sigma$
flux detection, corresponding to $AB(K)=24.8\pm 0.7$, or a 2-$\sigma$
upper limit of $AB(K)=24.5$.

The host galaxy is not detected in the $r$- and $I$-band.  We measure
2-$\sigma$ upper limits of $AB(r)=27.0$ and $AB(I)=27.0$.  The host is
not detected in the 850~$\mu$m image.  Based on the deboosted
850~$\mu$m fluxes of sub-mm sources observed in SCUBA images of
comparable r.m.s.\ noises (Coppin \etal\ 2006), we estimate a
4-$\sigma$ upper limit to the 850 $\mu$m flux of 7.6 mJy for the GRB
host.


\section{THE HOST GALAXY OF GRB\,080607}

To constrain the stellar population and star formation history of the
host galaxy, we consider a suite of synthetic stellar population
models generated using a revised Bruzual \& Charlot (2003) spectral
library that includes a new prescription for the TP-AGB evolution of
low- and intermediate-mass stars (Marigo \& Girardi 2007).  To improve
the uncertainties in the model analysis (such as the age-metallicity
degeneracy, e.g.\ Worthey 1994), we take into account known dust
properties from studies of early-time afterglow light (Perley \etal\
2010) and known ISM metallicity from absorption-line spectroscopy
(Prochaska \etal\ 2009).

The SED of the host galaxy is found to be extremely red, consistent
with the large extinction seen in the afterglow light.  As
demonstrated in Perley \etal\ (2010), the observed SED of the GRB
afterglow is best-described by a combination of a single power-law
spectrum (characteristic of the intrinsic afterglow radiation) and an
extinction law that is characterized by the broad 2175-\AA\ absorption
band, commonly seen in the Milky Way (MW) and the Large Magellanic
Cloud (LMC), and a large extinction column of $A_V\approx 3.1$ at
$z=3$.  However, the best-fit extinction curve appears to be
relatively flat at UV wavelengths with $R_V\approx 4$, and the
2175-\AA\ absorption feature is not as strong as what is seen in MW or
LMC.  Consequently, Perley \etal\ (2010) derived a best-fit extinction
curve for the host using a general extinction law from Fitzpatrick \&
Massa (1990), which allows variations in the UV slopes, and the depth
and width of the 2175-\AA\ absorption.

The large $R_V$ suggests that the GRB occurred in a dense environment
(e.g.\ Valencic \etal\ 2004), consistent with the large gas column
seen in afterglow spectra.  The extremely red colors of the host
galaxy suggest that the large amount of dust and gas surface mass
density seen in the afterglow light is not merely local to the line of
sight to the burst, but likely reflects the global dust content across
the entire host galaxy.
In addition, the presence of the 2175-\AA\ dust feature suggests that
the host galaxy resembles mature galaxies like MW or LMC, rather than
young star-forming systems like the Small Magellanic Cloud.  In the
following stellar population synthesis analysis, we include priors
from the best-fit Fitzpatrick \& Massa extinction law ($R_V=4.$ and
$A_V=3$) of Perley \etal\ (2010) and solar metallicity measured for
the host ISM from afterglow absorption-line observations.  We also
examine possible biases due to adopting these priors.

For the stellar population library, we adopt a Chabrier initial mass
function (IMF) with a range of star formation histories from a single
burst model to an exponentially declined star formation rate (SFR)
model with an e-folding time ranging from $\tau=0.1 - 2$~Gyr.  Over
the spectral range from rest-frame $\sim 1000$ \AA\ to $\sim 1$
micron, the observed emission is dominated by stellar light in dusty
galaxies (Hainline \etal\ 2009).  We therefore consider only stellar
components in the SED models.  We perform a maximum likelihood
analysis to compare the observed SED and a grid of model expectations,
taking into account both detections and non-detections.  The
likelihood function of this analysis is defined as
\begin{eqnarray}
{\cal L}(\tau,{\rm age}) = & &\left( \prod_{i=1}^{k} \exp \left\{ -\frac{1}{2} \left[ \frac{f_i -
\bar{f}(\tau,{\rm age})}{\sigma_i} \right]^2 \right\} \right) \times \nonumber \\
& & \left(\prod_{i=1}^l \int_{-\infty}^{f_i} df' \exp \left\{ -\frac{1}{2} \left[ 
\frac{f' - \bar{f}(\tau,{\rm age})}{\sigma_i} \right]^2 \right\} \right),
\end{eqnarray}
where $f_i$ is the observed flux (in $\mu$Jy) of the host in bandpass
$i$, $\bar{f}$ is the model expectation, and $\sigma_i$ is the
measurement uncertainty of $f_i$.  The first product of Equation (1)
extends over the $k$ measurements and the second product extends over
the $l$ upper limits.
The result shows that, with the adopted solar metallicity and a
relatively grey extinction law, the stellar population of the host
galaxy is best described by an exponentially declining SFR of $\tau=2$
at the age of $\sim 2.2$ Gyr.  The observed SED and the best-fit model
are presented in the top panel of Figure 2, showing a good agreement
in all bandpasses.  
The likelihood function of the stellar age is presented in the bottom
panel of Figure 2.

The best-fit stellar age of the host is similar to the age of the
universe at $z=3$, indicating that the host galaxy was formed very
early in time.  The comparable stellar age and star formation
e-folding time indicates that the host has been undergoing active star
formation since birth.  We estimate the SFR and total stellar mass of
the host galaxy based on extinction corrected UV luminosity and
mass-to-$B$-band light ratio.  The unobscured best-fit spectrum is
displayed in the top panel of Figure 2 (dash-dotted curve).  Adopting
the best-fit model, we derive a rest-frame, extinction corrected UV
luminosity of $L(1500)=1.8\times 10^{30}\ h^{-2}$ erg s$^{-1}$
Hz$^{-1}$ at 1500 \AA.  For a Chabrier IMF, this corresponds to ${\rm
SFR}\approx 125\,h^{-2}\,{\rm M}_\odot\,{\rm yr}^{-1}$ (Salim \etal\
2007).  At the age of $\approx 2$ Gyr, we derive a total stellar mass
of $M_*\sim 4\times 10^{11}\ h^{-2}\,{\rm M}_\odot$.  The uncertainty
in $M_*$ due to uncertainties in the star formation history is found
to be $\approx 50$\%.

\begin{figure}
\begin{center}
\includegraphics[scale=0.45]{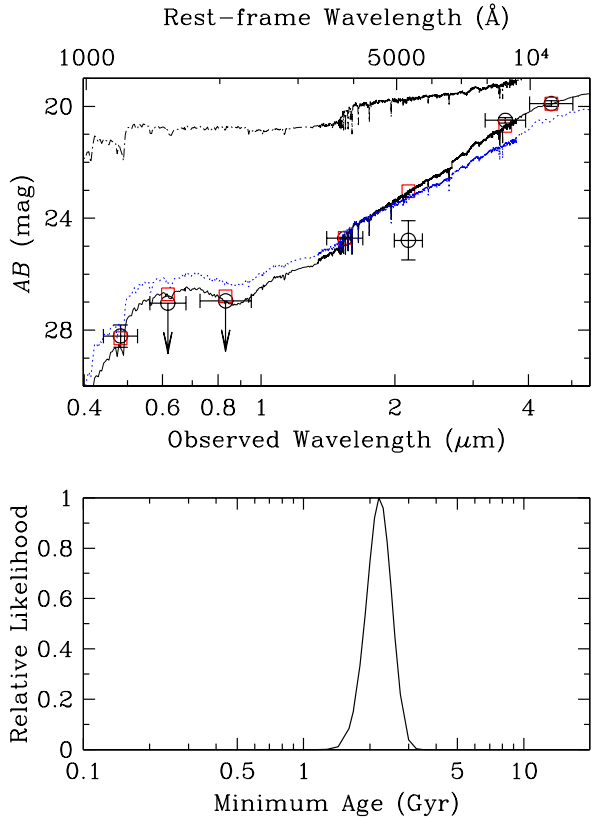}
\caption{Top: Comparison of the observed SED of the dusty host of
  GRB\,080607 at $z_{\rm GRB}=3.036$ and the best-fit stellar
  population synthesis model.  Optical and infrared photometric
  measurements are shown in open circles with errorbars.
  We place a 2-$\sigma$ upper limit to the observed brightness, when
  the galaxy is not detected.  The horizontal errorbars indicate the
  FWHM of each bandpass.  The solid curve represents the best-fit
  synthetic model after accounting for dust obscuration (the best-fit
  Fitzpatrick \& Massa law of Perley \etal\ 2010 with $R_V=4$ and
  $A_V=3$ at $z=3$).  The open squares represent the predicted
  brightness from the best-fit model.  The dash-dotted spectrum at the
  top shows the intrinsic spectrum prior to the application of dust
  obscuration.  The dotted curve represents a best-fit model based on
  the same extinction law but with $A_V=2.5$, indicating that reducing
  the amount of dust extinction no longer provides a good fit to the
  observed SED.  Adopting the 2-$\sigma$ upper limit for the $K$-band,
  $AB(K)>24.5$, does not change the results.
  Bottom: The likelihood function of the minimum age of the underlying
  stellar population as described by the broad-band SED.}
\end{center}
\end{figure}

To investigate possible biases due to adopting the afterglow
extinction curve and line-of-sight metallicity for the global
properties of the host galaxy, we repeat the stellar population
synthesis analysis with varying metallicity and the amount of dust
extinction.  We find that for a fixed amount of extinction, the
derived SFR and $M_*$ are insensitive to the adopted metallicity.
Reducing the amount of dust extinction in the model SED results in a
poor fit to the observed SED for any combination of star formation
history and metallicity (e.g.\ the dotted curve in Figure 2).  This
exercise confirms that the dust extinction law determined from the
afterglow light and the line-of-sight metallicity are representative
of the mean properties across the entire host galaxy.  We therefore
conclude that the best-fit SFR and $M_*$ are robust.

\section{DISCUSSION}

Our multi-wavelength imaging follow-up has uncovered an extremely red
galaxy at the location of the ``dark'' GRB\,080607 at $z_{\rm
GRB}=3.036$.  Given the coincident position and consistent red optical
and near-infrared colors, we argue that the galaxy is the host of the
dusty burst.  The host galaxy is clearly extended in the HST WF3/IR
F160W image with a half-light radius of $\approx 0.3''$, corresponding
to a physical half-light radius of $1.8\ h^{-1}$ kpc at $z=3$.  This
is typical of what is seen for star-forming galaxies at $z\sim 3$
(e.g.\ Bouwens \etal\ 2004).
The host galaxy is both massive and actively forming stars.  We
estimate the mean radiation field $I_0$ in the host ISM at near-UV
wavelengths ($\approx 1500-2000$ \AA) by averaging the
extinction-corrected UV flux over the extent of the host galaxy seen
in the HST WF3/IR image, and find that $I_0\approx 3\times 10^{-4}$
photons cm$^{-2}$ s$^{-2}$ Hz$^{-2}$.  The estimated mass and SFR are
among the highest known in the GRB host galaxy population (e.g.\
Christensen \etal\ 2004; Savaglio \etal\ 2009; Chen \etal\ 2009) and
in star-forming galaxies at $z\sim3$ (e.g.\ Erb \etal\ 2006), but are
comparable to the average of sub-mm sources (e.g.\ Borys \etal\ 2005;
Dye \etal\ 2006; Serjeant \etal\ 2008) and more than five times lower
than the brightest sub-mm galaxies (e.g.\ Tacconi \etal\ 2006).


Unobscured GRB host galaxies at intermediate redshifts appear to be
underluminous and low mass systems (e.g.\ Fruchter \etal\ 2006).  It
has therefore been argued that long-duration GRBs originate
preferentially in relatively metal deficient star-forming regions
(Wolf \& Podsiadlowski 2007; Modjaz \etal\ 2008). A low-metallicity
environment is favored by popular progenitor models so that the
progenitor star can preserve high spin and a massive stellar core to
produce a GRB (e.g.\ Yoon \& Langer 2005; Woosley \& Heger 2006).
However, statistical studies have shown that the distributions of ISM
metallicity and UV luminosity of known GRB host galaxies at $z>2$ are
consistent with the expectations of a UV luminosity selected
star-forming galaxy sample (Fynbo \etal\ 2008; Chen \etal\ 2009).

GRB080607 represents a unique example of a dark GRB that was luminous
enough to allow detailed observations of its afterglow despite it
occurring in a heavily obscured galaxy.  Thus we were able to secure
an unambiguous redshift of the dusty burst through afterglow
absorption-line spectroscopy and to identify the host galaxy based on
an astrometric match to the afterglow position.  Our current limit at
850$\mu$m is typical of what is seen in GRB host galaxies (e.g.\
Tanvir \etal\ 2004).  Improved sensitivity limit at the sub-mm
wavelengths will provide a different estimate of the SFR (e.g.\
Chapman \etal\ 2005) and some constraint on the dust-to-stellar mass
ratio of the host.  The large gas and dust mass uncovered in the
afterglow spectrum together with the IR bright host galaxy already
show that mature, dusty star-forming galaxies do contribute to the GRB
host galaxy population (see also Levesque \etal\ 2010), supporting the
notion that long-duration GRBs trace the bulk of cosmic star
formation.  Follow-up near-IR spectroscopy of the host will not only
confirm the host identification, but also allow a detailed study of
the gas kinematics in the host ISM.

\acknowledgments

We thank John Mulchaey for obtaining the IMACS $r$ images presented in
this paper and St\'ephane Charlot for providing the updated spectral
library.  Support for the HST DD program \#12005 was provided by NASA
through a grant from the Space Telescope Science Institute.
H.-W.C. acknowledges partial support from an NSF grant AST-0607510.
The James Clerk Maxwell Telescope is operated by The Joint Astronomy
Centre on behalf of the Science and Technology Facilities Council of
the United Kingdom, the Netherlands Organisation for Scientific
Research, and the National Research Council of Canada. 


\begin{references}

\vskip 0.2in

\reference{} Borys, C., Smail, I., Chapman, S. C., Blain, A. W.,
Alexander, D. M., \& Ivison, R. J. 2005, ApJ, 635, 853

\reference{} Bouwens, R. J., Illionworth, G. D., Blakeslee, J. P.,
Broadhurst, T. J., \& Franx, M. 2004, ApJ, 611, L1

\reference{} Bruzual, A. G. \& Charlot S. 2003, \mnras, 344, 1000

\reference{} Cenko, S. B. et al.\ 2009, ApJ, 693, 1484

\reference{} Chapman, S. \etal\ 2005, ApJ, 622, 772

\reference{} Chen, H.-W. \etal\ 2009, ApJ, 691, 152

\reference{} Chen, H.-W., Helsby, J. E., Gauthier, J.-R., Shectman,
S. A., Thompson, I. B., \& Tinker, J. L. 2010, ApJ, 714, 1521

\reference{} Christensen, L., Hjorth, J., \& Gorosabel, J. 2004., A\&A, 425, 913

\reference{} Coppin, K. \etal\ 2006, MNRAS, 372, 1621

\reference{} Dempsey, J. T., Friberg, P., Jenness, T., Bintley, D., \&
Holland, W. S. 2010, Proc. SPIE, 7741, 54

\reference{} Dressler, A., Hare, T., Bigelow, B. C., \& Osip, D. J. 2006, Proc. SPIE, 6269, 13

\reference{} Dye, S. \etal\ 2008, MNRAS, 386, 1107


\reference{} Erb, D. K., Steidel, C. C., Shapley, A. E., Pettini, M.,
Reddy, N. A., \& Adelberger, K. L. 2006, ApJ, 646, 107

\reference{} Faber, S. M. \etal\ 2007, ApJ, 665, 265

\reference{} Fitzpatrick, E. L. \& Massa, D. 1990, ApJS, 72, 163

\reference{} Fruchter, A. \etal\ 2006, Nature, 441, 463

\reference{} Fynbo, J. P. U. \etal\ 2006, A\&A, 451, L47

\reference{} Fynbo, J. P. U., Prochaska, J. X., Sommer-Larsen, J.,
Dessauges-Zavadsky, M., \& M{\o}ller, P. 2008, ApJ, 683, 321

\reference{} Greiner, J. \etal\ 2009, ApJ, 693, 1610

\reference{} Hainline, L. J., Blain, A. W., Smail, I., Frayer, D. T.,
Chapman, S. C., Ivison, R. J., \& Alexander, D. M. 2009, ApJ, 699,
1610


\reference{} Holland, W. \etal\ 2006,  Proc. SPIE, 6275, 45

\reference{} Jakobsson, P., Hjorth, J., Fynbo, J. P. U., Watson, D.,
Pedersen, K., Bj\"ornsson, G., \& Gorosabel, J. 2004, ApJ, 617, L21

\reference{} Kawai, N. et al.\ 2006, Nature, 440, 184.

\reference{} Levesque, E. M., Kewley, L. J., Graham, J. F., \&
Fruchter, A. S. 2010, ApJ, 712, L26

\reference{} Mangano, V. \etal\ 2008, GCN Circ.\ 7847

\reference{} Mannucci, F. et al.\ 2009, MNRAS, 398, 1915

\reference{} Marigo, P. \& Girardi, L. 2007, A\&A, 469, 239

\reference{} Modjaz, M. \etal\ 2008, AJ, 135, 1136

\reference{} Noterdaeme, P., Ledoux, C., Srianand, R., Petitjean, P.,
\& Lopez, S. 2009, A\&A, 503, 765

\reference{} Oke, J. B. et al.\ 1995, PASP, 107, 375

\reference{} Perley, D. A. \etal\ 2009, AJ, 138, 1690

\reference{} Perley, D. A. \etal\ 2010, AJ submitted (arXiv:1009.0004)


\reference{} Prochaska, J.~X., Chen, H.-W., Wolfe, A. M.,
Dessauges-Zavadsky, M., \& Bloom, J. S. 2008, ApJ, 672, 59

\reference{} Prochaska, J.~X. \etal\ 2009, ApJ, 691, L27

\reference{} Salvaterra, R. \etal\ 2009, Nature, 461, 1258

\reference{} Savaglio, S. 2006, New Journal of Physics, 8, 195

\reference{} Savaglio, S., Glazebrook, K., \& Le Borgne, D. 2009, ApJ,
691, 182

\reference{} Serjeant, S. \etal\ 2008, MNRAS, 386, 1907

\reference{} Sheffer, Y., Prochaska, J. X., Draine, B. T., Perley,
D. A., \& Bloom, J. S. 2009, ApJ, 701, L63

\reference{} Sonnentrucker, P., Welty, D. E., Thorburn, J. A., \&
York, D. G. 2007, ApJS, 168, 58

\reference{} Srianand, R., Noterdaeme, P., Ledoux, C., \& Petitjean,
P. 2008, A\&A, 482, L39

\reference{} Tacconi, L.J. \etal\ 2006, ApJ, 640, 228

\reference{} Tanvir, N. R. \etal\ 2004, MNRAS, 352, 1073

\reference{} Tanvir, N. R. \etal\ 2009, Nature, 461, 1254

\reference{} Valencic, L. A., Clayton, G. C., \& Gordon, K. D. 2004, ApJ, 616, 912

\reference{} Wolf, C. \& Podsiadlowski, P. 2007, MNRAS, 375, 1049

\reference{} Worthey, G. 1994, ApJS, 95, 107

\reference{} Woosley, S. E. \& Heger, A. 2006, ApJ, 637, 914

\reference{} Yoon, S.-C. \& Langer, N. 2005, A\&A, 443, 643

\end{references}

\clearpage

\begin{center}
{\bf Erratum: ``A Mature Dusty Star-forming Galaxy Hosting GRB\,080607 at $z=3.036$'' (2010, ApJ, 723, L218)}
\end{center}

We have discovered an error in the photometric measurements of the
host galaxy in our Spitzer IRAC images.  The host is detected in the
IRAC 3.5$\mu$m and 4.5$\mu$m channels with $AB({\rm 3.5}\mu)=22.9\pm
0.2$ and $AB({\rm 4.5}\mu)=22.7\pm 0.2$ mag.  The photometric
measurements of the host galaxy in other bandpasses remain unchanged.
Adopting the revised Spitzer IRAC photometry and the original optical
and near-IR photometric measurements, we estimate the total stellar
mass ($M_*$) and on-going star formation rate (SFR) of the host galaxy
based on the stellar population synthesis analysis described in Chen
\etal\ (2010).  Given the uncertainties in the global dust content of
the host galaxy, we allow the metallicity, $A_V$, and dust extinction
law to differ from what were found in the afterglow light (e.g.\
Prochaska \etal\ 2009; Perley \etal\ 2010).  The likelihood analysis
described in Equation (1) of Chen \etal\ (2010) yields an
extinction-corrected SFR of $(8-12)\ h^{-2}\,{\rm M}_{\odot}\,{\rm
  yr}^{-1}$, a mean ISM radiation field $I_0\approx (2.3-3.5)\times
10^{-5}$ photons cm$^{-2}$ s$^{-2}$ Hz$^{-1}$, and $M_*=(0.5-1.4)
\times 10^{10}\ h^{-2}\,{\rm M}_{\odot}$ for the host galaxy.  These
are about an order of magnitude lower than those originally published.
We note that the uncertainties in the derived $M_*$, $I_0$ and SFR are
driven by the uncertainties in the global dust extinction law of the
host galaxy.  The galaxy is still fairly massive, but not as extreme
as previously thought.  The observed spectral energy distribution
(SED) of the host galaxy is presented in the revised figure below,
together with the best-fit synthetic model of super-solar metallicity
and a Milky-Way type dust extinction law of $A_V=1.25$.  A Fitzpatrick
\& Massa (FM) law described in Perley \etal\ (2010) with $A_V=1.8$ and
super-solar metallicity produces a similarly good fit to the observed
SED.

Adopting the priors from the best-fit dust obscuration, the
Fitzpatrick \& Massa (FM) law from Perley \etal\ (2010) with $R_V=4.2$
and $A_V=3.3$ at $z=3$ and solar metallicity from Prochaska \etal\
(2009) leads to a similar estimate of $M_*$ ($\sim 5\times 10^9\
h^{-2}\,{\rm M}_\odot$) but significantly higher SFR ($\sim 230\
h^{-2}\,{\rm M}_{\odot}\,{\rm yr}^{-1}$).  However, this model also
predicts an observed optical brightness that is $\Delta\,AB=1$ mag
brighter than the observed 2-$\sigma$ upper limits in the $r$ and $I$
bands.  We therefore consider this model unlikely to represent the
global extinction property of the host galaxy.  The result indicates a
large spatial variation in the dust content across the host galaxy.

\begin{figure}[hb]
\begin{center}
\includegraphics[scale=0.45]{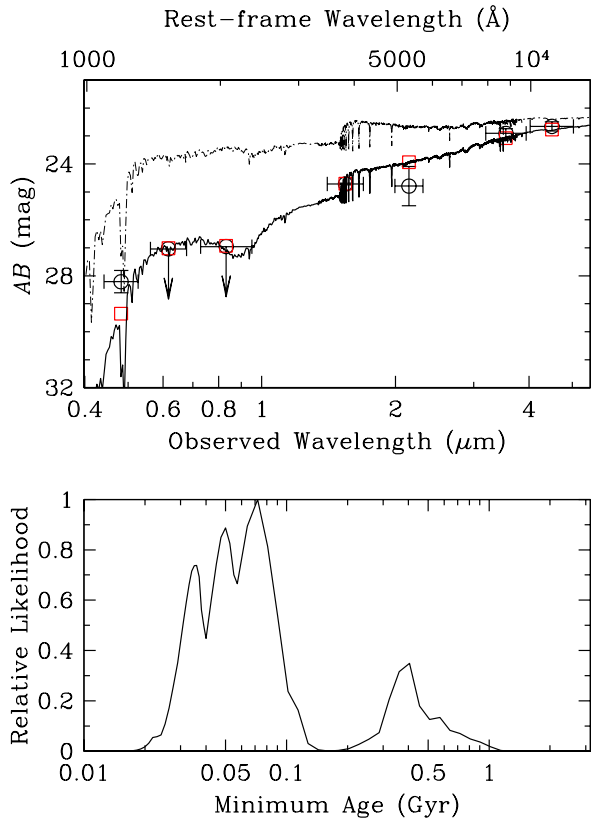}
\caption{Top: Comparison of the observed SED of the dusty host of
  GRB\,080607 at $z_{\rm GRB}=3.036$ and the best-fit stellar
  population synthesis models.  Optical and infrared photometric
  measurements are shown in open circles with errorbars.  We place a
  2-$\sigma$ upper limit to the observed brightness, when the galaxy
  is not detected.  The horizontal errorbars indicate the FWHM of each
  bandpass.  The solid curve represents the best-fit synthetic model
  of super-solar metallicity and a Milky-Way type dust extinction law
  of $A_V=1.25$.  The open squares represent the predicted brightness
  from this best-fit model.  The thin dash-dotted spectrum at the top
  shows the intrinsic spectrum prior to the application of dust
  obscuration.  Bottom: The likelihood functions of the minimum age of
  the underlying stellar population as described by the broad-band
  SED.}
\end{center}
\end{figure}

\end{document}